\documentclass[12pt]{iopart}

\usepackage{graphicx}

\begin{document}

\title{Low-frequency anomalies in dynamic localization}

\author{Stefano Longhi}

\address{Dipartimento di Fisica,
Politecnico di Milano and Istituto di Fotonica e Nanotecnologie del Consiglio Nazionale delle Ricerche, Piazza L. da Vinci 32, I-20133 Milano, Italy}
\ead{longhi@fisi.polimi.it}
\begin{abstract}
Quantum mechanical spreading of a particle hopping on  tight binding lattices can be suppressed by the application of an external ac force, leading to periodic wave packet reconstruction. Such a phenomenon, referred to as dynamic localization (DL), occurs for certain magic values of the ratio $\Gamma=F_0/ \omega$ between the amplitude $F_0$ and frequency $\omega$ of the ac force. It is generally believed that in the low-frequency limit ($\omega \rightarrow 0$) DL can be achieved for an infinitesimally small value of the force $F_0$, i.e. at finite values of $\Gamma$. Such a normal behavior is found in homogeneous lattices as well as in inhomogeneous lattices of Glauber-Fock type. Here we introduce a tight-binding lattice model with inhomogeneous hopping rates, referred to as pseudo Glauber-Fock lattice,  which shows DL but fails to reproduce the normal low-frequency behavior of homogeneous and Glauber-Fock lattices. In pseudo Glauber-Fock lattices, DL can be exactly realized, however at the DL condition the force amplitude $F_0$ remains finite as $\omega \rightarrow 0$. Such an anomalous behavior is  explained in terms of a $\mathcal{PT}$ symmetry breaking transition of an associated two-level non-Hermitian Hamiltonian that effectively describes the dynamics of the Hermitian lattice model.
\end{abstract}




\maketitle

\section{Introduction}

Quantum mechanical diffusion of a particle hopping on a tight binding lattice can be suppressed by the application of an external ac force, which leads to periodic wave packet reconstruction \cite{Dunlap}. Such a phenomenon, referred to as dynamic localization (DL), was originally predicted in a seminal paper by Dunlap and Kenkre \cite{Dunlap} for a particle hopping on a homogeneous lattice and explained as a collapse of the quasi energy spectrum at certain 'magic' driving amplitudes \cite{Holthaus}.  Several works have extended the theory of DL to different physical conditions, for example to account for disorder \cite{disorder}, non-nearest neighbor hopping \cite{sterco}, 
 particle interaction \cite{inter}, lattice truncation \cite{Holthaus,LonghiPRB2008,Longhi2013}, multiple ac fields or combined ac and dc fields \cite{combined,referee1}, and non-Hermitian lattices \cite{LonghiNH}. In particular, the addition of an ac field to a dc field generally destroys particle re-localization, however if the ratio between the ac modulation frequency and the Bloch oscillation frequency is a rational
number, a form of dynamic localization, so-called quasi-Bloch
oscillations, occurs irrespective of the shape and amplitude
of the ac component of the field \cite{referee1}.
 On the experimental side, DL has been observed in coherent lattice transport in different physical systems, including 
electronic transport in semiconductor superlattices \cite{semi}, 
 cold atoms and Bose-Einstein condensates in optical lattices \cite{BEC} and optical beams in curved waveguide arrays \cite{optics0,optics,optics2}.  
In a homogeneous lattice, for a sinusoidal force $F=F_0 \cos (\omega t)$ with amplitude $F_0$ and frequency $\omega$,  assuming $\hbar=1$ and a lattice period $a=1$, the DL condition is obtained when the normalized amplitude $\Gamma=F_0/ \omega$ of the force is a root of the $J_0$ Bessel function, i.e. when \cite{Dunlap,Holthaus}
\begin{equation}
J_0(\Gamma)=0.  
\end{equation}
Such a result is valid in the nearest-neighbor approximation, that we will assume in all the following analysis. Remarkably, the condition of DL as given by Eq.(1) holds also for the so-called class of Glauber-Fock lattices \cite{Longhi2013}, which show an inhomogeneous hopping rate \cite{note}. An implication of Eq.(1) is that, in the low frequency limit $\omega \rightarrow 0$, DL can be achieved for an infinitesimally small value of the forcing amplitude, i.e.
\begin{equation}
\frac{F_0}{\omega}= \Gamma
\end{equation}
remains finite as $\omega \rightarrow 0$. Such a condition will be referred to as the 'normal' behavior of DL in the low-frequency limit. \par
 In this work we present an ac-driven lattice model with inhomogeneous hopping rates, referred to as pseudo Galuber-Fock lattice, which shows exact DL but fails to follow the normal behavior  of Eq.(2). In particular, at the DL point $F_0/\omega$ diverges like $1 / \omega$ as $\omega \rightarrow 0$. Such an anomalous behavior can be explained in terms of a $\mathcal{PT}$ symmetry breaking transition of an associated two-level non-Hermitian Hamiltonian that effectively describes the dynamics of the pseudo Glauber-Fock lattice model.

\section{Lattice models}
Let us consider the hopping motion of a quantum particle on a tight-binding lattice driven by an external ac force $F(t)$ of frequency $\omega$. Assuming a lattice period $a=1$ and $\hbar=1$, in the nearest-neighbor approximation the amplitude probability $c_n(t)$ to find the particle at lattice site $n$ is described by the set of coupled equations \cite{Dunlap}
\begin{equation}
i \frac{dc_n}{dt}= -\kappa_{n}c_{n-1}-\kappa_{n+1} c_{n+1}+nF(t) c_n
\end{equation}
where $\kappa_n$ is the hopping rate between sites $n$ and $(n-1)$. The regime of DL corresponds to periodic self-reconstruction of an arbitrary initial wave packet at integer multiplies of the forcing period $T=2 \pi / \omega$, i.e.
\begin{equation}
|c_n(lT)|^2=|c_n(0)|^2
\end{equation}
for $l=1,2,3,...$ and for an arbitrary initial state $c_n(0)$  For a homogeneous lattice $\kappa_n=\kappa$ it is well-known that DL is attained provided that the following condition is satisfied \cite{Dunlap}
\begin{equation}
\int_0^T dt \exp \left[  i \int_0^t dt' F(t') \right]=0.
\end{equation}
For example, for a sinusoidal force $F(t)=F_0 \cos (\omega t)$ the condition (5) yields
\begin{equation}
J_0 ( \Gamma)=0
\end{equation}
where $\Gamma= F_0 / \omega$ and $J_0$ is the Bessel function of first kind and zero order. Taking the lowest root $\Gamma \simeq 2.405$ of the Bessel function,
it follows that DL is achieved for a force amplitude $F_0$ given by
\begin{equation}
F_0 \simeq 2.405 \omega.
\end{equation}
Such a result indicates that, in the low frequency limit $\omega \rightarrow 0$, DL can be achieved for an infinitesimally small value of the force amplitude $F_0$. The possibility to realize DL with an infinitesimally small amplitude of the forcing in the low modulation frequency limit will be referred to as  the {\it normal} behavior of DL at low frequencies.
Of course the normal behavior of DL holds for other modulation profiles of the force, for example for a square-wave modulation. Remarkably, in a recent work \cite{Longhi2013} it was shown that DL can be exactly realized also for certain inhomogeneous semi-infinite lattices, the class of so-called Glauber-Fock (GF) lattices \cite{SzamGF,SzamGFb}. In such lattices, the hopping rate is inhomogeneous and increases with the site number $n$ according to the relation
\begin{equation}
\kappa_n =\sigma \sqrt{n},
\end{equation}
($n=0,1,2, 3, ...$).  Noticeable, lattices with a square-root dependence on the site index were earlier introduced in Ref.\cite{referee2} to study the equivalence between continuous and discrete Schr\"{o}dinger equations for a quantum particle in a dc field by preserving the Heisenberg equations of motion.
As shown in Ref.\cite{Longhi2013}, in such lattices DL can be realized under the same condition Eq.(5)  of a homogeneous lattice. Such a result indicates that  the normal regime of DL in the low frequency limit found in homogeneous lattices holds for the GF lattices as well.\par 
Let us now consider another class of semi-infinite lattices with inhomogeneous hopping rates given by
\begin{equation}
\kappa_n =\sigma n.
\end{equation}
Such a class of lattices, that will be referred to as pseudo GF lattices, differs for the GF lattice  [Eq.(8)] because of a {\it linear} (rather than {\it square root}) increase of the hopping rate with the site index $n$. Such a class of inhomogeneous lattices was previously introduced in Ref.\cite{LonghiBO} and shown to admit of exact Bloch oscillations in presence of a dc force. Here we will investigate the case of an ac force. As shown in the next sections, DL can be realized for the pseudo GF lattices as well, however as opposed to the homogeneous or GF lattices they show an {\it anomalous}   behavior in the low frequency limit.

\section{Dynamic localization condition}
An important property of both GF and pseudo GF lattices is that the single-particle hopping dynamics on the lattices, as described by Eq.(3), can be mapped onto the dynamics  in Fock space of certain bosonic field models. For the GF lattices,  let us consider a bosonic field described by the time-periodic Hamiltonian \cite{Longhi2013,SzamGF,SzamGFb}
\begin{equation}
\hat{H}(t)=F(t) \hat{a}^{\dag} \hat{a}-\sigma(\hat{a}+\hat{a}^{\dag})
\end{equation}
 where $\hat{a}^{\dag}$ and $\hat{a}$ are the creation and destruction operators of the bosonic field, satisfying the usual communtation relations $[\hat{a},\hat{a}]=[\hat{a}^{\dag},\hat{a}^{\dag}]=0$ and $[\hat{a},\hat{a}^{\dag}]=1$. If we expand the state vector $| \psi(t) \rangle$ of the system in Fock space as 
 \begin{equation}
 |\psi(t) \rangle=\sum_{n=0}^{\infty}c_n(t) |n \rangle
 \end{equation}
  where $|n \rangle=(1 / \sqrt{n !}) \hat{a}^{\dag \; n} |0 \rangle$, it can be readily shown that the evolution of the amplitude probabilities $c_n(t)$ is governed by Eq.(3) with $\kappa_n$ given by Eq.(8) (GF lattices). Similarly, let us consider two bosonic fields described by the Hamiltonian
 \begin{equation}
\hat{H}(t)=\frac{F(t)}{2} (\hat{a}^{\dag} \hat{a}+\hat{b}^{\dag} \hat{b}) -\sigma(\hat{a}\hat{b}+\hat{b}^{\dag} \hat{a}^{\dag})
\end{equation}
where $\hat{a}^{\dag}$, $\hat{a}$ and $\hat{b}^{\dag}$, $\hat{b}$ are the creation and destruction operators of the two bosonic fields.  If we only consider the dynamics of the bosonic fields in the sub-space of Fock space corresponding to the states $|n \rangle=(1/ n!) \hat{a}^{\dag n} \hat{b^{\dag n}} |0 \rangle$ with the {\it same} boson number $n$ in the two fields, i.e. if we expand the state vector  $|\psi(t) \rangle$ of the fields as in Eq.(11), it readily follows that the evolution of the amplitude probabilities $c_n(t)$ is governed by Eq.(3) with $\kappa_n$ given by Eq.(9) (pseudo GF lattices). Hence the dynamical properties of the GF and pseudo GF lattices can be derived from the analysis of the Hamiltonians (10) and (12), which are quadratic in the bosonic field operators. Interestingly, the conditions of DL for the two lattice models can be obtained without the need to derive the explicit forms of the propagators for the two lattice systems. To this aim, let us notice that the site occupation probabilities $|c_n(t)|^2$ can be derived from the relation
\begin{equation}
|c_n(t)|^2=\frac{1}{2 \pi} \int_{-\pi}^{\pi} dq S(q,t) \exp(-iqn)
\end{equation}
where the spectrum $S(q,t)$ is given by
\begin{equation}
S(q,t)=\sum_{n=0}^{\infty} |c_n(t)|^2 \exp(iqn).
\end{equation}
From Eqs.(13) and (14) it follows that the condition for DL [Eq.(4)] can be written as
\begin{equation}
S(q,lT)=S(q,0)
\end{equation}
$(l=1,2,3,...$).
On the other hand, the spectrum $S(q,t)$ can be determined as the expectation value of the operator $\exp(iq \hat{a}^{\dag} \hat{a})$, i.e.
\begin{equation}
S(q,t)= \langle \psi(t) | \exp(iq \hat{a}^{\dag} \hat{a}) \psi(t) \rangle
\end{equation}
where $|\psi(t) \rangle$ is the state vector of the bosonic fields, with  $|n \rangle=(1 / \sqrt{n !}) \hat{a}^{\dag \; n} |0 \rangle$ for the GF lattice and  $|n \rangle=(1 / n!) \hat{a}^{\dag \; n} \hat{b}^{\dag \; n} |0 \rangle$ for the pseudo GF lattice. Let us indicate by $\hat{U}(t)$ the propagator associated to the time-periodic Hamiltonian $\hat{H}(t)$, i.e. $i  (d \hat{U}/dt)=\hat{H}(t) \hat{U}(t)$ and $| \psi(t) \rangle= \hat{U}(t) | \psi(0) \rangle$, and let us indicate by $\hat{A}_h(t)=\hat{U}^{\dag}(t) \hat{A} \hat{U}(t)$ the Heisenberg operator associated to any given operator $\hat{A}$ acting on the bosonic fields with $\hat{A}_h(0)=\hat{A}$. For a time-independent operator $\hat{A}$, the operator $\hat{A}_h(t)$ then satisfies the Heisenberg equation of motion
\begin{equation}
i \frac{d \hat{A}_h}{dt}= \left[ \hat{A}_h, \hat{H}_h \right].
\end{equation}
In the Heisenberg representation, the spectrum $S(q,t)$ as given by Eq.(16) can be computed according to the relation
\begin{equation}
S(q,t)= \langle \psi(0) | \exp \left\{  iq \hat{a}^{\dag}_{h}(t) \hat{a}_h(t) \right\} \psi(0) \rangle
\end{equation}
and hence the condition (15) for DL is realized provided that 
\begin{equation}
\hat{a}_h(T)=\hat{a}_{h}(0) \exp(i \varphi)=\hat{a} \exp(i \varphi) 
\end{equation}
where $\varphi$ is an arbitrary phase constant.\\
Let us first consider the DL problem for the GF lattice, which was previously studied in Ref.\cite{Longhi2013}. In this case, from Eq.(17) with $\hat{A}=\hat{a}$ one obtains
\begin{equation}
i \frac{d \hat{a}_h}{dt}=F(t) \hat{a}_h-\sigma
\end{equation}
which is readily solved with the initial condition $\hat{a}_h(0)=\hat{a}$ yielding
\begin{equation}
\hat{a}_h(t)=\exp \left[-i \int_0^t dt'F(t')\right] \left\{ \hat{a}+i \sigma \int_0^t dt' \exp \left[ i \int_0^{t'} d \xi F( \xi) \right] \right\}.
\end{equation}
Note that, if the condition (5) is satisfied, taking into account that $\int_0^T dt F(t)=0$ one has $\hat{a}_{h}(T)=\hat{a}$. This shows that, as previously proven in Ref.\cite{Longhi2013}, DL occurs in GF lattices  for the same driving conditions as in the homogenous lattices.\par
Let us now consider the pseudo GF lattice. In this case, coupled equations for the operators $\hat{a}_h, \hat{b}^{\dag}_h$ (and similarly for  $\hat{b}_h, \hat{a}^{\dag}_h$) can be derived from Eq.(17) by taking $\hat{A_h}=\hat{a}_h$ and $\hat{A}_h=\hat{b}^{\dag}_{h}$. One obtains
\begin{equation}
i \frac{d}{dt} \left(
\begin{array}{c}
\hat{a}_h \\
\hat{b}^{\dag}_{h}
\end{array}
\right)= \mathcal{M}(t) \left(
\begin{array}{c}
\hat{a}_h \\
\hat{b}^{\dag}_{h}
\end{array}
\right)
\end{equation}
where the $ 2 \times 2$ matrix $\mathcal{M}(t)$ is given by
\begin{equation}
\mathcal{M}(t)=
\left(
\begin{array}{cc}
F(t)/2 & - \sigma \\
\sigma & - F(t)/2
\end{array}
\right).
\end{equation}
Note that the matrix $\mathcal{M}(t)$ is {\it non-Hermitian}, in spite the original Hamiltonian $\hat{H}(t)$ of the pseudo GF lattice [Eq.(12)] is Hermitian. Since $\mathcal{M}(t)$ is periodic with period $T= 2 \pi / \omega$, Floquet theory applies. Indicating by $\mathcal{U}$ the monodromy matrix of the linear periodic system (22), i.e. the propagator from $t=0$ to $t=T$
\begin{equation}
\left(
\begin{array}{c}
\hat{a}_h(T) \\
\hat{b}^{\dag}_{h}(T)
\end{array}
\right)= \mathcal{U} \left(
\begin{array}{c}
\hat{a} \\
\hat{b}^{\dag}
\end{array}
\right)
\end{equation}
Floquet theory ensures that $\mathcal{U}$ can be written in the form
\begin{equation}
\mathcal{U}=\mathcal{F} \exp(-i \mathcal{J}T) \mathcal{F}^{-1}
\end{equation}
where $\mathcal{F}$ is a non-singular $2 \times 2$ matrix and $\mathcal{J}$ is a Jordan normal form. The elements (eigenvalues) $\mu_1$ and $\mu_2$ on the diagonal of $\mathcal{J}$ are the Floquet exponents (quasi energies) of the periodic system and 
are defined to within integer multiplies of $ \omega$.  Here we will assume for the sake of definiteness that the real parts of $\mu_{1,2}$ are inside the interval $(-\omega/2, \omega/2)$. Hence the DL condition [Eq.(19)] for the pseudo GF lattice requires 
\begin{equation}
\mu_1=\mu_2.
\end{equation}
The quasi-energies (Floquet exponents) $\mu_1$ and $\mu_2$, and thus the DL points corresponding to quasi-energy crossing, can be numerically computed by solving the system (22) over one oscillation cycle using standard methods. Owing to the form of the matrix $\mathcal{M}(t)$, it turns out that $\mu_2=-\mu_1$. Note that, since $\mathcal{M}(t)$ is non-Hermitian, the quasi energies are generally complex numbers.
\begin{figure}[htb]
\centerline{\includegraphics[width=10cm]{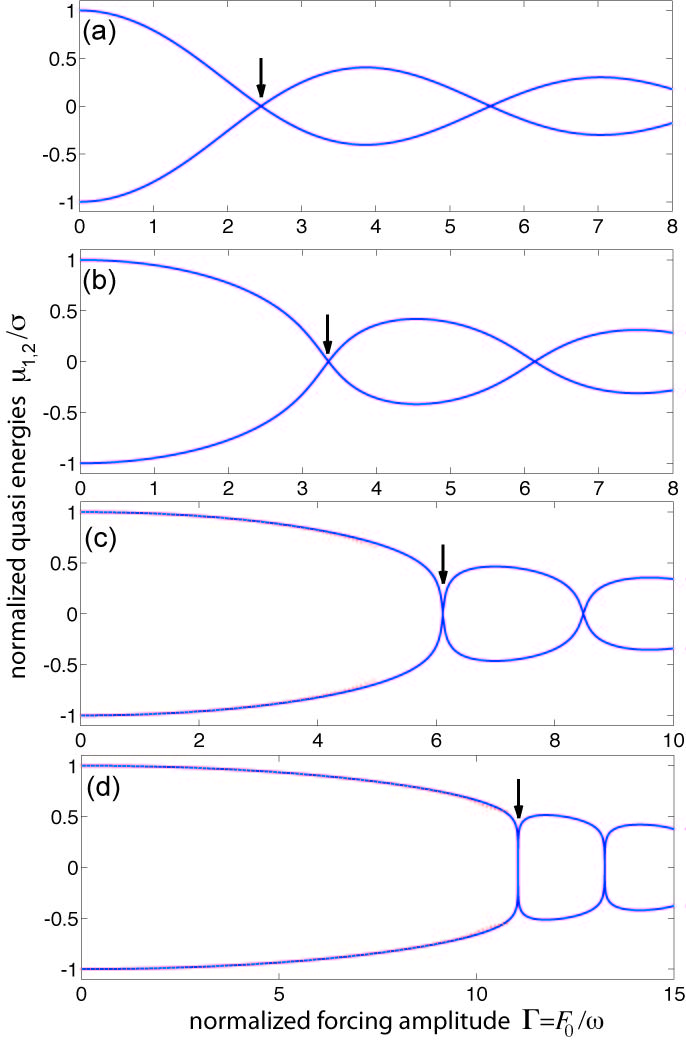}} \caption{
Numerically-computed behavior of the quasi energies $\mu_{1,2}$ (imaginary parts) versus $\Gamma=F_0 / \omega$ for the non-Hermitian two-level system (22) with a sinusoidal driving force $F(t)=F_0 \cos (\omega t)$ for decreasing values of $\omega / \sigma$: (a) $\omega / \sigma=5$, (b) $\omega / \sigma=1$, (c) $\omega / \sigma=0.4$ and (d) $\omega / \sigma=0.2$. The vertical arrows in the panels show the lowest value $\Gamma=\Gamma_0$ at which level crossing, corresponding to DL, is found. The thin dotted curves in (c) and (d), almost overlapped with the solid ones, show the behavior of quasi energies as predicted by the WKB analysis [Eq.(27)]. The quasi energies in the WKB limit are plotted for forcing amplitudes $F_0$ smaller than $2 \sigma$ to avoid turning points. }
\end{figure}

\section{Low-frequency anomalies of DL in the pseudo Glauber-Fock lattices}
An important property of pseudo GF lattices is that DL can never be realized in the low frequency limit $\omega \rightarrow 0$ by a forcing amplitude $F_0$ which is vanishing like $\sim \omega$, i.e. pseudo GF lattices fail to show the normal low-frequency  behavior of DL of homogeneous and GF lattices.  To prove this statement, let us notice that in the low-frequency limit $\omega \rightarrow 0$ the quasi-energies $\mu_{1,2}$ can be asymptotically determined by a WKB analysis of Eq.(22), yielding
\begin{equation}
\mu_{1,2}= \pm \frac{1}{T} \int_0^T dt \sqrt {\frac{F^2(t)}{4}-\sigma^2}.
\end{equation}
This expression of the quasi energies is accurate provided that no turning points arise over the oscillation cycle. If the forcing amplitude is small enough such that $|F(t)| < 2 \sigma$ over the entire oscillation cycle, there are no turning points and the quasi-energies, as given by Eq.(27), are manifestly imaginary and complex conjugate each other, so as energy crossing $\mu_1=\mu_2$ requested for DL can not be realized. Hence DL in pseudo GF lattices shows an anomalous behavior as $\omega \rightarrow 0$. As an example, in Fig.1 we show the numerically-computed behavior of the quasi-energies versus $\Gamma=F_0 / \omega$ for a sinusoidal forcing $F(t)=F_0 \cos (\omega t)$ and for decreasing values of the modulation frequency $\omega$, normalized to $\sigma$. The quasi energies turn out to be imaginary, so that in the figure the imaginary parts of $\mu_1$ and $\mu_2$ are shown. DL is obtained for the normalized forcing amplitudes $\Gamma$ where the quasi energies cross. The lowest value $\Gamma=\Gamma_0$ of level crossing is indicated by an arrow in Fig.1. Note that, as $\omega / \sigma$ gets small [from Fig.1(a) to Fig.1(d)], the amplitude $\Gamma_0$ at the lowest quasi energy crossing gets large \cite{note2}, indicating the anomalous low-frequency behavior of DL for the pseudo GL lattices. Conversely, in the large frequency modulation regime DL is attained at values of $\Gamma$ that are (as expected) close to the roots  $J_0(\Gamma)=0$ [see Fig.1(a)]. In Fig.2 we also show, as an example, the self-imaging property of the pseudo GF lattice in the DL regime corresponding to single site excitation of the lattice at initial time.\par
Failure of the normal behavior of DL in the low frequency limit  can be elegantly explained on the basis of the {\it non-Hermitian} nature of the Heisenberg equations (22-23) describing the evolution of the bosonic field operators $\hat{a}_h(t)$ and $\hat{b}^{\dag}_h(t)$.  Following Ref.\cite{Dunlap},  for a sinusoidal force $F(t)=F_0 \cos (\omega t)$ in the limit $\omega \rightarrow 0$ and for a force amplitude $F_0$ that is finite or is vanishing but of {\it lower} order than $\omega$, for times $t \ll T$ the particle dynamics on the lattice can be described by replacing the ac force with the dc force $F=F_0$. In this limit, for 
a homogeneous lattice it is well known that the particle undergoes Bloch oscillations with periodicity $T_B=2 \pi / F_0$, with no restrictions on the smallness of the force $F_0$. Such a result holds for the GF lattice as well, as it can be readily proven from Eq.(21). Conversely, for the pseudo GF lattice taking a dc force $F(t)=F_0$ from Eqs.(22) and (23) one obtains 
\begin{figure}[htb]
\centerline{\includegraphics[width=12cm]{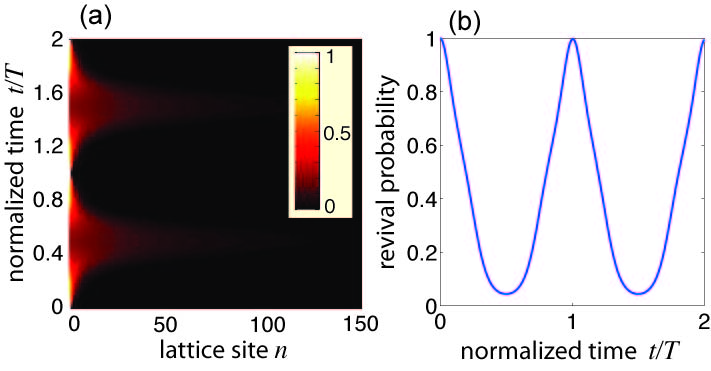}} \caption{
(a) Numerically-computed evolution of the amplitude probabilities (snapshot of $|c_n(t)|$) in a sinusoidally-driven pseudo GF lattice for parameter values $\sigma=1$, $\omega=1$ and $\Gamma=F_0/ \omega=3.353$, corresponding to DL [see Fig.1(b)]. The initial condition corresponds to single site excitation of the edge site, i.e. $c_n(0)=\delta_{n,0}$. In (b) the evolution of the revival probability $|c_0(t)|^2$ is also shown. Time is normalized to the period $T=2 \pi / \omega$ of the ac force.}
\end{figure}
\begin{equation}
\hat{a}_h(t)=\left\{\cos(\lambda t)-i \frac{F_0}{2 \lambda} \sin(\lambda t) \right\} \hat{a}+i \frac{\sigma}{\lambda} \sin (\lambda t) \hat{b}^{\dag}
\end{equation}
where
\begin{equation}
\lambda=\sqrt{(F_0/2)^2-\sigma^2}
\end{equation}  
and $\pm \lambda$ are the two eigenvalues of the non-Hermitian matrix $\mathcal{M}$ for $F(t)=F_0$. From Eq.(28) it follows that, if the eigenvalues $ \pm \lambda$ are complex, i.e. for $F_0< 2 \sigma$, the self-imaging condition $\hat{a}_h(t)=\hat{a} \exp(i \varphi)$ is not satisfied at any time $t$, whereas when the eigenvalues $\pm \lambda$ are real, i.e. for $F_0> 2 \sigma$, self imaging, corresponding to Bloch oscillations, are observed with the periodicity $T_B= \pi/ \lambda$, i.e.
\begin{equation}
T_B= \frac{ \pi}{\sqrt{(F_0/2)^2-\sigma^2}}
\end{equation}
which agrees with the result previously reported in Ref.\cite{LonghiBO}.
Hence in the pseudo GF lattices self-imaging is found when the eigenvalues of the non-Hermitian matrix $\mathcal{M}$ become real, which requires a {\it minimum} value $F_0=2 \sigma$ of the forcing amplitude. Below such a value, i.e. for $F_0<2 \sigma$, the eigenvalues of the matrix $\mathcal{M}$ are complex conjugates and self-imaging can not be realized. Hence the occurrence of self-imaging in the pseudo GF lattice can be traced back to a $\mathcal{PT}$ symmetry breaking transition \cite{Bender} at $F_0=2 \sigma$ of the non-Hermitian system describing the evolution of the bosonic operators in the Heisenberg picture. For a dc field, the $\mathcal{PT}$ symmetry breaking transition at $F_0= 2 \sigma$ is also associated to a {\it metal-insulator phase transition} of the eigenstates of the pseudo GF lattice Hamiltonian $\hat{H}$, which can be calculated in a closed form as shown in Ref. \cite{LonghiBO}: while all the eigenstates of $\hat{H}$ are extended for $F_0 < 2 \sigma$ (continuous spectrum), they become all localized for $F_0> 2 \sigma$ (point spectrum). 
 
\section{Conclusions}
Quantum diffusion of a particle hopping on a tight-binding lattice is known to be suppressed by
application of an ac force. Such a phenomenon, referred to as dynamic localization, corresponds to periodic wave packet re-localization induced the ac force at certain magic amplitudes, and it is thus conceptually very different from other forms of localization, like Anderson localization. Recent experimental observations of DL for matter and classical waves \cite{BEC,optics} have renewed the interest in such a rather old problem. One of the general belief of DL is that in the low-frequency limit  suppression of quantum diffusion can be achieved for an infinitesimally small value of the force amplitude. This normal behavior occurs in homogeneous lattices \cite{Dunlap,Holthaus} and in other integrable inhomogeneous lattice models which are known to show exact DL, like Glauber-Fock lattices  \cite{Longhi2013}. In this work we have introduced a tight-binding lattice model with inhomogeneous hopping rates, referred to as pseudo Glauber-Fock lattice,  in which  DL is exact but shows an anomalous behavior in the low-frequency limit. In particular,  at the DL point the force amplitude $F_0$ should remain finite, regardless of the smallness of the modulation frequency $\omega$. Such an anomalous behavior has been explained in terms of a $\mathcal{PT}$ symmetry breaking transition \cite{Bender} of an associated two-level non-Hermitian Hamiltonian, that effectively describes the dynamics of the Hermitian lattice model. In the limit of a dc force, the $\mathcal{PT}$ symmetry breaking transition of the non-Hermitian two-level system is associated to a metal-insulator phase transition of the lattice eigenstates from extended to localized. Light transport in sinusoidally-curved optical waveguide lattices with engineered hopping rates can provide a possible physical system for the experimental observation of low-frequency anomalies of DL \cite{optics,optics2}. In such a system, the hopping rates can be tailored by a suitable control of the waveguide separation, whereas a sinusoidal ac field can be mimicked by sinusoidally curving the waveguide axis along the propagation direction;  a possible scheme is the zig-zag array geometry discussed in Ref.\cite{LonghiBO}. To observe the anomalous behavior of DL in the low frequency regime, one can design different sets of waveguide arrays corresponding to a decreasing value of $\omega/ \sigma$; for each set, output light intensity distributions can be measured, after one period of the oscillation cycle, for single-site input excitation at different values of the bending amplitude (see, for instance, Fig.6 of the experiment discussed in Ref.\cite{optics0}). The low-frequency anomaly of DL can be thus demonstrated by the increase of $\Gamma$ at the DL point when $\omega / \sigma$ is decreased. An experimentally challenging issue is the ability to precisely control the tunneling rates over several waveguides, thus realizing arrays with a sufficiently number of waveguides that reproduce the square-root law of hopping rates and avoid truncation effects at the right edge.  In the zig-zag geometry discussed in Ref.\cite{LonghiBO}, a maximum number of waveguides of $\sim 70$  is expected to be feasible, which is enough to show the increase of $\Gamma$ from $\sim 2.405$ in the high $\omega / \sigma$ regime [see Fig.1(a)] to $\sim 3.353$ at $\omega / \sigma=1$ [see Figs.1(b) and 2].\\ 
Our results provide novel insights into the phenomenon of dynamic localization and show a noteworthy example where a symmetry-breaking transition in a non-Hermitian model associated to an Hermitian system can reveal a qualitative change of the energy spectrum and particle dynamics \cite{note3}.


\begin{thebibliography}{10}
\bibitem{note}
For a lattice with arbitrary inhomogeneous hopping rates,  DL can be only approximately obtained in the high-frequency regime under the condition (1), whereas in the low-frequency regime DL can not be realized.
\bibitem{note2}
For the homogeneous and GF lattices $\Gamma_0$ does not depend on $\omega$ and is equal to 2.405.
\end{thebibliography}
\end{document}